\documentclass[8.5pt,twoside,twocolumn]{article}
\usepackage[super,sort&compress,comma]{natbib}
\usepackage{mhchem}
\usepackage{times,mathptmx}
 \usepackage{times}
\usepackage{sectsty}
\usepackage{balance}
\usepackage{hyperref}
\usepackage{epsfig}
\usepackage{amsfonts}
\usepackage{amssymb}
\usepackage{gensymb}
\usepackage{graphicx}
\usepackage{dcolumn}
\usepackage{bm}
\usepackage[latin1]{inputenc}
\usepackage{color}
\usepackage{enumerate}

\def \degree {^\mathrm{o}}

\newcommand{\etal}[1]{\textit{et al.}}

\renewcommand{\thefootnote}{\alph{footnote}}

\usepackage{lastpage}
\usepackage[format=plain,justification=raggedright,singlelinecheck=false,font=small,labelfont=bf,labelsep=space]{caption}
\usepackage{fancyhdr}

\begin{document}

\twocolumn[
  \begin{@twocolumnfalse}
\noindent\LARGE{\textbf{Strong dynamical effects during stick-slip adhesive peeling}}
\vspace{0.6cm}

\noindent\large{\textbf{Marie-Julie Dalbe,\footnotemark[1]\footnotemark[2] St\'{e}phane
Santucci,\footnotemark[1] Pierre-Philippe Cortet,\footnotemark[3] and Lo\"{i}c Vanel\footnotemark[2]\let\thefootnote\relax\footnotemark [4]}}\vspace{0.5cm}

\vspace{0.2cm}

\noindent\normalsize(Dated: November 14, 2013)
\vspace{0.6cm}

\noindent \normalsize{We consider the classical problem of the
stick-slip dynamics observed when peeling a roller adhesive tape
at a constant velocity. From fast imaging recordings, we extract
the dependencies of the stick and slip phases durations with the
imposed peeling velocity and peeled ribbon length.
Predictions of Maugis and Barquins [in \textit{Adhesion 12},
edited by K.W. Allen, Elsevier ASP, London, 1988, pp. 205--222]
based on a quasistatic assumption succeed to describe
quantitatively our measurements of the stick phase duration. Such
model however fails to predict the full stick-slip cycle duration,
revealing strong dynamical effects during the slip phase.}
\vspace{0.5cm}

 \end{@twocolumnfalse}
  ]
\footnotetext[1]{\textit{Laboratoire de Physique de l'ENS Lyon, CNRS and Universit\'{e} de Lyon, France}}
\footnotetext[2]{\textit{Institut Lumi\`ere Mati\`ere, UMR5306 Universit\'e Lyon 1-CNRS, Universit\'e de Lyon, France.}}
\footnotetext[3]{\textit{Laboratoire FAST, CNRS, Univ. Paris Sud, France.}}
\let\thefootnote\relax\footnotetext[4]{\textit{M.-J. Dalbe, E-mail : mariejulie.dalbe@ens-lyon.fr; S. Santucci, E-mail : stephane.santucci@ens-lyon.fr; P.-P. Cortet, E-mail : ppcortet@fast.u-psud.fr; L. Vanel, E-mail : loic.vanel@univ-lyon1.fr}}

\section{Introduction}

Everyday examples of adhesive peeling are found in applications
such as labels, stamps, tape rollers, self-adhesive envelops or
post-it notes. During the peeling of those adhesives, a dynamic
instability of the fracture process corresponding to a jerky
advance of the peeling front and called ``stick-slip'' may occur.
This stick-slip instability has been an industrial concern since
the 1950's because it leads to noise levels above the limits set
by work regulations, to adhesive layer damage and/or to mechanical
problems on assembly lines. Nowadays this instability is still a
limiting factor for industrial productivity due to the limitations
of generic technical solutions applied to suppress it, such as
anti-adhesive silicon coating.

From a fundamental point of view, the stick-slip instability of
adhesive peeling is generally understood as the consequence of an
anomalous decrease of the fracture energy $\Gamma (v_p)$ of the
adhesive-substrate joint in a specific range of peeling front
velocity
$v_p$.~\cite{Gardon1963,Gent1969,Aubrey1969,Barquins1986,Maugis1988,Derail1997,Barquins1997,Aubrey1980}
Indeed, when the peeling process also involves a compliance
between the point where the peeling velocity is imposed and the
fracture front, this decreasing fracture energy naturally leads to
oscillations of the fracture velocity $v_p$ around the mean
velocity $V$ imposed by the operator. Often, it is simply the
peeled ribbon elasticity which provides a compliance to the
system. From a microscopic perspective, such anomalous decrease of
the fracture energy $\Gamma(v_p)$ (correctly defined for stable
peeling only) could correspond (but not necessarily) to transition
from cohesive to adhesive failure~\cite{Gent1969,Aubrey1969} or
between two different interfacial failure
modes.~\cite{Derail1997,Aubrey1980} More fundamentally, this
decrease of the fracture energy has been proposed to be the
consequence of the viscous dissipation in the adhesive
material.~\cite{Maugis1985} De Gennes~\cite{degennes1996} further
pointed out the probable fundamental role of the adhesive material
confinement (which was evidenced experimentally in
ref.~\citenum{Aubrey1969}) in such viscoelastic theory. Since
then, it has however appeared that a model based on linear
viscoelasticity solely cannot be satisfactory and that the role of
creep, large deformations and temperature gradient in the adhesive
material is important
(refs.~\citenum{Gent1996,Carbone2005,Barthel2009,Tabuteau2011} and
references therein).

Experimentally, the stick-slip instability was first characterized
thanks to peeling force measurements which revealed strong
fluctuations in certain ranges of peeling
velocity.~\cite{Gardon1963,Aubrey1969,Barquins1986, Maugis1988}
Since then, it has also been studied through indirect measurements
of the periodic marks left on the
tape~\cite{Racich1975,Barquins1986,Ryschenkow1996, Maugis1988} or
of the emitted acoustic noise.~\cite{Gandur1997,Ciccotti2004}
Thanks to the progress in high speed imaging, it is now possible
to directly access the peeling fracture dynamics in the stick-slip
regime.~\cite{Cortet2007,Thoroddsen2010,Cortet2013}

In the late 1980's, Barquins and co-workers,~\cite{Barquins1986,
Maugis1988} performed a series of peeling experiments of a
commercial adhesive tape (3M
Scotch\textsuperscript{\textregistered} 602) at constant pulling
velocity $V$ and for various lengths of peeled ribbon $L$. For the
considered adhesive, the velocity range for which stick-slip was
evidenced, thanks to peeling force fluctuations measurements, was
shown to be $0.06<V<2.1$~m~s$^{-1}$. In a subrange of unstable
peeling velocity $0.06<V<0.65$~m~s$^{-1}$, the authors succeeded
to access the stick-slip cycle duration thanks to the post-mortem
detection of periodic marks left on the tape by stick-slip events.
Moreover, they managed to model quantitatively the measured
stick-slip period,~\cite{Barquins1986, Maugis1988} assuming the
fracture dynamics to remain a quasistatic problem during the stick
phase and backing on measurements of the stable branch of the
fracture energy $\Gamma(v_p)$ at low peeling velocities below the
instability onset.

In this article, we revisit these experiments by studying the
stick-slip dynamics during the peeling of a roller adhesive tape
at an imposed velocity. The principal improvement compared to
Barquins's seminal work is that, thanks to a high speed camera
coupled to image processing, we are able to access the dynamics of
the peeling fracture front. We focus on the study of the duration
of the stick-slip cycle and its decomposition into stick and slip
events, which data are inaccessible through other techniques. We
present experimental data of the stick and slip durations for a
wide range of imposed peeling velocity $V$ and for different
peeled ribbon lengths $L$. We show that the model proposed by
Barquins and co-workers~\cite{Barquins1986,Maugis1988} describes
the evolution of the duration of the stick phase, but fails to
predict the duration of the whole stick-slip cycle due to
unexpectedly long slip durations.

\section{Experimental setup}

\begin{figure}
\centerline{\includegraphics[width=8.6cm]{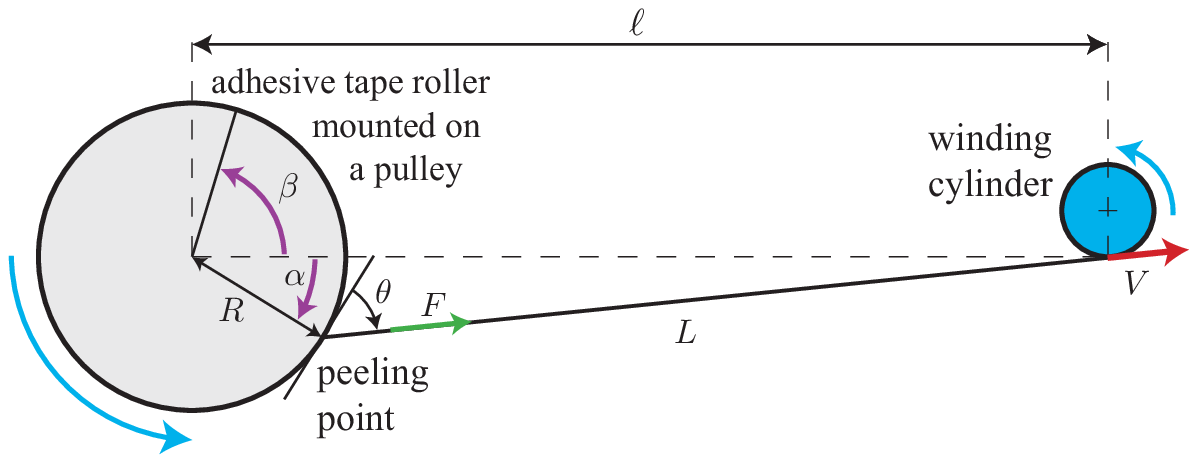}}
    \caption{\label{fig:manip}(Color online) Schematic view of the
    experimental setup. The angles $\alpha$ and $\beta$ are oriented
    \emph{clockwise} and \emph{counterclockwise} respectively. Roller
    diameter: $40$~mm $<2R<58$~mm, roller and tape width: $b=19$~mm.}
\end{figure}

In this section, we describe briefly the experimental setup which
has already been presented in details in a recent
work.~\cite{Cortet2013} We peel an adhesive tape roller (3M
Scotch\textsuperscript{\textregistered} $600$, made of a
polyolefin blend backing coated with a layer of a synthetic
acrylic adhesive, also studied in
refs.~\citenum{Cortet2007,Cortet2013,Barquins1997}), mounted on a
freely rotating pulley, by winding up the peeled ribbon extremity
on a cylinder at a constant velocity $V$ using a servo-controlled
brushless motor (see Fig.~\ref{fig:manip}). The experiments have
been performed at a temperature of $23\pm 2\degree$ and a relative
humidity of $45\pm 5$\%. The width of the tape is $b=19$~mm, its
thickness $e=38~\mu$m and its Young modulus $E=1.26$~GPa.

Each experiment consists in increasing the winding velocity from
$0$ up to the target velocity $V$. Once the velocity $V$ is
reached, it is maintained constant during two seconds before
decelerating the velocity back to zero. When stick-slip is present
this 2-second stationary regime of peeling provides sufficient
statistics to compute well converged stick-slip mean features. We
have varied the imposed velocity $V$ from $0.0015$ to
$2.5$~m~s$^{-1}$ for different values of the peeled tape length
between $L=0.08$ and $1.31$~m. During an experiment, the peeled
tape length $L$ (Fig.~\ref{fig:manip}) is submitted to variations,
due to the stick-slip fluctuations and to slow oscillations of the
peeling point angular position, which however always remain
negligible compared to its mean value (less than
0.3\%).~\cite{Cortet2013}

\section{Peeling force measurement}

Thanks to a force sensor
(Interface\textsuperscript{\textregistered} SML-5) on the holder
maintaining the pulley, we are able to measure the mean value of
the force $F$ transmitted along the peeled tape during one
experiment. When peeling is stable, we compute the strain energy
release rate $G$ from the mean value of the force $F$, following
the traditional relation for the peeling
geometry~\cite{Rivlin1944,Kendall1975}
\begin{equation}
    G=\frac{F}{b}(1-\cos \theta)+\frac{1}{2Ee}\left( \frac{F}{b}\right)^2 \simeq \frac{F}{b},\label{eq.G}
\end{equation}
for a peeling angle $\theta \simeq 90 \degree$ (see
Fig.~\ref{fig:manip}). The quantity $G$ corresponds to the amount
of mechanical energy released by the growth of the fracture by a
unit surface. The right-hand term of eqn~(\ref{eq.G}) finally
simply takes into account the work done by the operator but
discards the changes in the elastic energy stored in material
strains (term $(F/b)^2/2Ee$ in
eqn~(\ref{eq.G}))~\cite{Kendall1975} which are negligible here.
Indeed, the maximum encountered force in our experiments is
typically of about $2$~N, which gives $F/b\approx 100$~J~m$^{-2}$,
to be compared to $(F/b)^2/2Ee\approx 0.12$~J~m$^{-2}$.

In the context of elastic fracture mechanics, the condition for a
fracture advance at a constant velocity $v_p$ is a balance between
the release rate $G$ and a fracture energy $\Gamma(v_p)$ required
to peel a unit surface and accounting for the energy dissipation
near the fracture front. When the fracture velocity $v_p$
approaches the Rayleigh wave velocity $v_R$, $\Gamma(v_p)$ also
takes into account the kinetic energy stored in material motions
which leads to a divergence when $v_p \rightarrow
v_R$.~\cite{Freund1998} In our system, the strain energy release
rate $G$, computed through eqn~(\ref{eq.G}), therefore stands as a
measure of the fracture energy $\Gamma(v_p)$ when the peeling is
stable only, \textit{i.e.} when $v_p$ is constant. We will
nevertheless compute $G$ for the experiments in the stick-slip
regime for which the peeling fracture velocity $v_p(t)$ is
strongly fluctuating in time. In such a case, $G$ cannot be used
as a measure of a fracture energy: it is simply the time average
of the peeling force $F$ in units of $G$.

\begin{figure}
\centerline{\includegraphics[width=8.7cm]{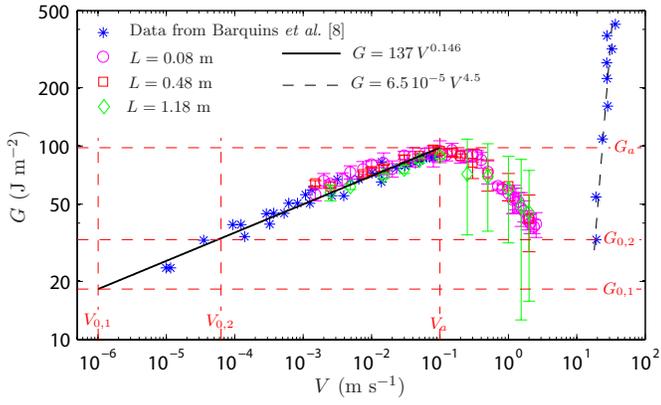}}
    \caption{\label{fig:force}(Color online) Mean value of the peeling
    force $F$, in units of strain energy release rate $G=F/b$, as a
    function of $V$ for 3 different peeled tape lengths $L$. Stars
    report the data of Barquins and Ciccotti~\cite{Barquins1997} for
    the same adhesive. Solid line is a power law fit
    $G=137\,V^{0.146}$ of the data in the low velocity stable branch.
    Errorbars represent the standard deviation of the force
    fluctuations during one experiment.}
\end{figure}

In Fig.~\ref{fig:force}, we plot $G$ as a function of the imposed
peeling velocity $V$ for three different peeled tape lengths $L$.
When the peeling is stable, the peeling force is nearly constant
in time, whereas it fluctuates strongly when stick-slip
instability is present. The standard deviation of these
fluctuations is represented in Fig.~\ref{fig:force} with error
bars. Large error bars are indicative of the presence of
stick-slip.

Between $V=0.0015$~m~s$^{-1}$ and $V=0.10 \pm 0.03$~m~s$^{-1}$, we
observe that $G=F/b$ increases slowly with $V$ and that its
temporal fluctuations are nearly zero, revealing that the peeling
is stable. This increasing branch $G(V)$ is therefore a measure of
the adhesive fracture energy $\Gamma(v_p=V)=G(V)$ for $V<0.10 \pm
0.03 $~m~s$^{-1}$. Our results are compatible with the data
reported by Barquins and Ciccotti~\cite{Barquins1997} for the same
adhesive tape (see Fig.~\ref{fig:force}). However, they  explored
a much larger range of velocities in this stable branch of
peeling, down to $V=10^{-5}$~m~s$^{-1}$. Using both series of
measurements, it is reasonable to model the stable peeling branch
with a power law, $G(V)=a\,V^{n}$, with $n=0.146$ and $a=137$. For
$0.10\pm 0.03$~m~s$^{-1}<V \leqslant 2.5$~m~s$^{-1}$, we observe
that the measured value of $G(V)$ decreases with $V$. This
tendency, which was already observed in previous
experiments,~\cite{Yamazaki2006} is accompanied with the
appearance of temporal fluctuations which are the trace of the
stick-slip instability. From these data, we can estimate the onset
of the instability to be $V_a=0.10\pm 0.03$~m~s$^{-1}$. The
measured decreasing branch of $G(V)$ for $V>V_a$ appears as a
direct consequence of the anomalous decrease of the fracture
energy at the origin of the instability. It is important to note
that the measured mean value of $G=F/b$ is nearly independent of
the length of peeled ribbon $L$. This result is natural in the
stable peeling regime but was a priori unknown in the stick-slip
regime.

Barquins and Ciccotti~\cite{Barquins1997} succeeded to measure a
second stable peeling branch for $V\geqslant 19$~m~s$^{-1}$. This
increasing branch constitutes a measure of the peeling fracture
energy $\Gamma(v_p=V)=G(V)$ in a fast and stable peeling regime.
In ref.~\citenum{Barquins1997}, this branch is inferred to exist
for velocities even lower than $V=19$~m~s$^{-1}$, although it was
not possible to measure it. Backing on the data of
ref.~\citenum{Maugis1988} for a very close adhesive, one can
however guess that the local minimum value of $G(V)$,
corresponding to a velocity in the range
$2.5$~m~s$^{-1}<V<19$~m~s$^{-1}$, would be bounded by
$G_{0,1}=18<G<G_{0,2}=33$~J~m$^{-2}$.

\section{Peeling point dynamics}

The local dynamics of the peeling point is imaged using a high
speed camera (Photron FASTCAM SA4) at a rate of $20\,000$~fps. The
recording of each movie is triggered once the peeling has reached
a constant average velocity $V$ ensuring that only the stationary
regime of the stick-slip is studied. Through direct image
analysis,~\cite{Cortet2013} the movies allow access to the
curvilinear position of the peeling point
$\ell_{\alpha}(t)=R\,\alpha$ in the laboratory frame (with
$\alpha$ the angular position of the peeling point and $R$ the
roller diameter, $\alpha >0$ in Fig.~\ref{fig:manip}). Image
correlations on the adhesive tape roller contrast pattern further allow direct access to its angular velocity $d\beta/dt(t)$ in the
laboratory frame (where $\beta$ is the unwrapped angular position
of the roller, $\beta >0$ in Fig.~\ref{fig:manip},
$\ell_{\beta}(t)=R\,\beta$). We finally compute numerically the
curvilinear position $\ell_p(t)=\ell_\beta(t)+\ell_\alpha(t)$ and
velocity $v_{p}(t)=d\ell_p/dt$ of the peeling point in the roller
reference frame.

The curvilinear position of the peeling point $\ell_{\alpha}(t)$
in the laboratory frame is actually estimated from the position of
the peeled ribbon at a small distance $0.30\pm 0.05$~mm from the
peeling fracture front on the roller surface. We therefore do not
detect strictly the peeling fracture front position but a very
close quantity only. This procedure can consequently introduce
some bias in our final estimation of the fracture front velocity
$v_p(t)$. This bias is notably caused by the changes in the radius of curvature of the tape at the junction with the substrate which are due to the force oscillations in the peeled tape characteristics of the stick-slip instability. Such effect actually biases the measurement toward larger velocities during the stick phase and lower velocities during the slip phase. Another effect that leads to uncertainties on velocity measurement is the emission of a transverse wave in the peeled tape when the fracture velocity abruptly changes at the beginning and at the end of slip phases.

\begin{figure}
\centerline{\includegraphics[width=8.6cm]{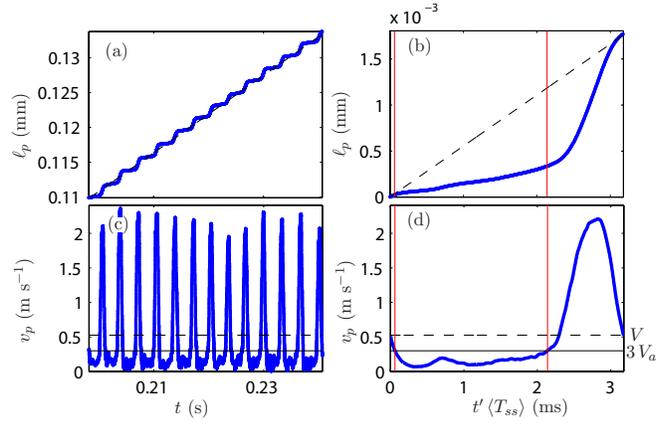}}
\caption{\label{fig:ppposition}(Color online) (a) Peeling point
position $\ell_p(t)$ in the roller reference frame for an
experiment performed at $V=0.55$~m~s$^{-1}$ and $L=0.47$~m. The
dashed line shows $\ell_p=V\,t$, with $V$ the average peeling
velocity. (b) Corresponding phase averaged peeling point position
as a function of $t' \, \langle T_{ss} \rangle$ (see main text).
(c) and (d) Corresponding instantaneous (c) and phase averaged (d)
peeling point velocity $v_p$. The dashed horizontal lines show the
average peeling velocity $V$ and the continuous horizontal lines show
$3\,V_a$. In (b) and (d), the vertical lines show the transitions
between the stick ($v_p<3\,V_a$) and the slip ($v_p>3\,V_a$)
phases.}
\end{figure}

Figs.~\ref{fig:ppposition}(a) and (c) represent the fracture
position $\ell_p(t)$ and velocity $v_p(t)$ as a function of time
for a typical experiment performed at $V=0.55$~m~s$^{-1}$ and
$L=0.47$~m. In these figures, we observe alternate phases of slow
--stick phase-- and fast --slip phase-- peeling which are the
signature of the stick-slip motion. These large velocity
fluctuations are quite regular in terms of duration and to a
lesser extent in terms of amplitude at least at the considered
peeling velocity. Our general data analysis further consists in
the decomposition of the signal of instantaneous peeling velocity
$v_{p}(t)$ into stick-slip cycles by setting the beginning of each
cycle at times $t_n$ ($n$ denoting the $n^{th}$ cycle) when
$v_p(t_n) = V$ and $dv_p/dt(t_n)<0$. From this data, we extract the
duration $T_{ss}$ of each stick-slip cycle for which we define a
rescaled time $t'=(t-t_n)/T_{ss}$. We further compute the phase
averaged evolution of the peeling fracture velocity $v_p(t')$ from
$t'=0$ to $1$ considering all the stick-slip cycles in one
experiment. With this procedure, we finally extract for each
peeling velocity $V$ and peeled tape length $L$ the typical
fracture velocity evolution during a stick-slip cycle getting rid
of intrinsic fluctuations of the stick-slip period. In
Figs.~\ref{fig:ppposition}(b) and (d), we show the phase averaged
position and velocity profiles, corresponding to Figs.~\ref{fig:ppposition}(a) and (c) respectively, as a function of $t=t'\,\langle
T_{ss} \rangle$ ($\langle
\, \rangle$ denotes the ensemble averaged value over all the
cycles in one experiment).

From these phase averaged velocity profiles, we define, for each
experimental condition $V$ and $L$, stick events as continuous
periods during which $v_{p}(t)<3\,V_a$ and slip events as
continuous periods during which $v_{p}(t)>3\,V_a$. According to
the model of Barquins \etal,~\cite{Barquins1986, Maugis1988} a
natural threshold in order to separate the stick and slip phases
is the onset of the instability $V_a$ (as defined in
Fig.~\ref{fig:force}). However, as discussed previously, due to the procedure used for the
detection of the peeling point, our measurement of the fracture
velocity can be affected by biases caused by the variation the tape curvature at the peeling point and by the propagation of transverse waves in the tape. The effect of the later can be observed in Fig.~\ref{fig:ppposition}(d) in the early stage of the stick phase. In order to avoid taking into account the velocity biases in the decomposition of the stick-slip cycle, we chose for the threshold separating the stick and slip phases a value little larger the ``theoretical'' threshold $V_a$, that is to say $3\,V_a$.

Finally, as we have shown recently in ref.~\citenum{Cortet2013},
when the peeling velocity $V$ is increased, low frequency pendular
oscillations of the peeling angle $\theta$ develop. Due to a
dependence of the stick-slip instability onset with the mean
peeling angle, these oscillations lead to intermittencies in the
stick-slip dynamics for peeling velocities $V>1.5$~m~s$^{-1}$. We
therefore exclude the experiments with $V>1.5$~m~s$^{-1}$ in the
sequel. For the studied experiments, we have a mean peeling angle
$\langle \theta \rangle = 90 \pm 3\degree$ with slow temporal
variations in the range $\Delta \theta = \pm 15 \degree$ during
one experiment.

\section{Stick-slip cycle duration}

\begin{figure}
\centerline{\includegraphics[width=7cm]{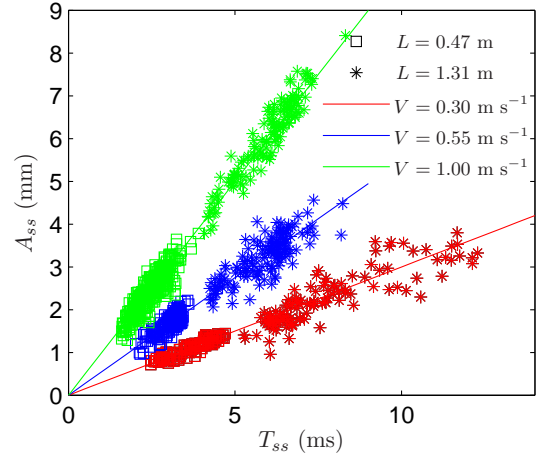}}
\caption{\label{AvsT}(Color online) Stick-slip amplitude $A_{ss}$
as a function of stick-slip period $T_{ss}$ for each
stick-slip cycle in 6 different experiments with $L=0.47$ and
$1.31$~m and $V=0.30$, $0.55$ and $1.00$~m~s$^{-1}$. The lines
represent the curves $A_{ss}=V\,T_{ss}$.}
\end{figure}

From the signal of peeling point position $\ell_p(t)$ (see
Fig.~\ref{fig:ppposition}(a)), we define the stick-slip amplitude
$A_{ss}$ as the distance travelled by the fracture during a
stick-slip cycle. In Fig.~\ref{AvsT}, we report this amplitude
$A_{ss}$ for each stick-slip event as a function of the
corresponding stick-slip period $T_{ss}$, for all events in 6
different experiments. These data gather close to the curve $A_{ss}
= V\,T_{ss}$. The large spread of the data along the curve $A_{ss}
= V\,T_{ss}$ reflects the statistics of the stick-slip cycle
amplitude and duration which could be due for instance to adhesive
heterogeneities. On the contrary, the dispersion of the data
around the curve $A_{ss} = V\,T_{ss}$ is much smaller. It actually
estimates the discrepancy between the imposed velocity $V$ and the
averaged fracture velocity for each stick-slip cycle. The observed
small discrepancy actually both traces back measurement errors on the
instantaneous fracture velocity and intrinsic fluctuations of the
dynamics.

In Fig.~\ref{AvsT}, one can already see that the statistically
averaged values of $A_{ss}$ and $T_{ss}$ increase with $L$ for a
given peeling velocity $V$. In the following, we will focus on the
study of the statistical average $\langle T_{ss} \rangle$ of the
duration of the stick-slip oscillation and its decomposition into
stick and slip phases with in mind the aim of testing the
description of Barquins, Maugis and
co-workers.~\cite{Barquins1986, Maugis1988} There is no need to
study the averaged stick-slip amplitude $\langle A_{ss} \rangle$
since it is univocally related to $\langle T_{ss} \rangle$ through
$\langle A_{ss} \rangle=V\,\langle T_{ss} \rangle$.

\begin{figure}
\centerline{\includegraphics[width=7cm]{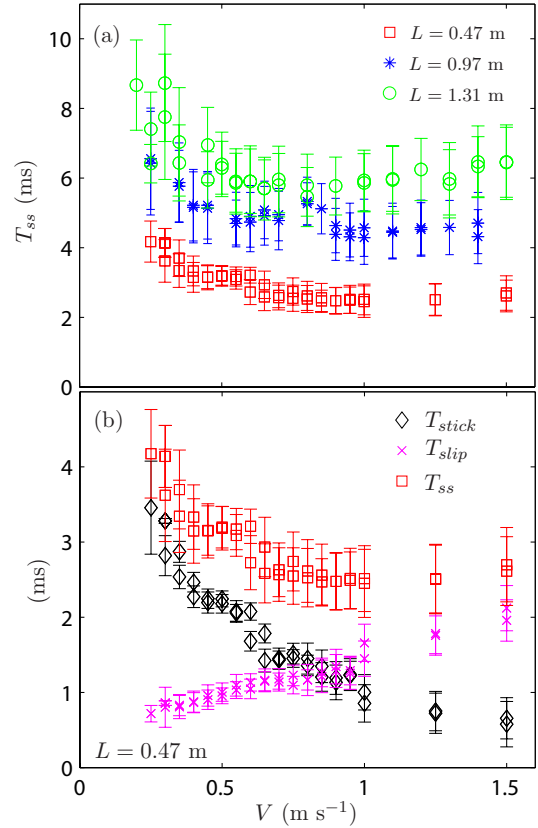}}
    \caption{\label{duree}(Color online) Average stick-slip cycle
    duration $T_{ss}$ as a function of the average peeling velocity
    $V$, for different lengths of the peeled ribbon $L$. (b) Average
    stick-slip, stick and slip durations as function of the average
    peeling velocity for $L=0.47$~m. Each data point corresponds to
    the average and each error bar to the standard deviation of the
    statistics over one experiment.}
\end{figure}

In Fig.~\ref{duree}(a), we plot the mean stick-slip duration
$T_{ss}$ as a function of $V$ for three different lengths $L$ of
the peeled ribbon. The data corresponds to the average $\langle
T_{ss}\rangle$ and the error bars to the standard deviation of the
statistics of $T_{ss}$ over all the stick-slip events in each
experiment. In the following, since we will consider the averaged
values only, we will skip the brackets $\langle \, \rangle$. At
first sight, it appears that, within the error bars, the stick-slip
duration $T_{ss}$ is stable over the major part of the explored
range of peeling velocity $V$. One can however note that,
independently of $L$, $T_{ss}$ tends to decrease with $V$ for $V
\leqslant V_c = 0.6 \pm 0.1$~m~s$^{-1}$. Such behavior is
compatible with the observations of Barquins \etal
,~\cite{Barquins1986} but appears here over a rather limited
velocity range. The characteristic velocity $V_c=0.6 \pm
0.1$~m~s$^{-1}$ above which $T_{ss}$ is nearly constant seems not
to depend strongly on the length of the peeled ribbon $L$.

\begin{figure}
\centerline{\includegraphics[width=7cm]{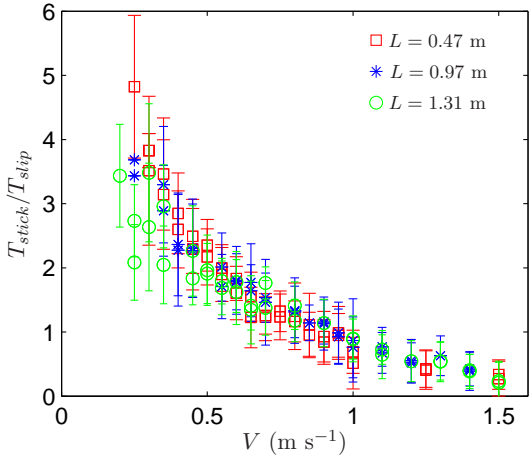}}
    \caption{\label{rapports}(Color online)  $T_{stick}/T_{slip}$
    \textit{vs.} $V$ for 3 different $L$. Each data point corresponds
    to the average and each error bar to the standard deviation of the
    statistics over one experiment.}
\end{figure}

In Fig.~\ref{duree}(b), we show the mean durations of stick and
slip events, $T_{stick}$ and $T_{slip}$ respectively, as a
function of the imposed peeling velocity $V$ for the experiments
performed with the peeled length $L=0.47$~m. Interestingly, we
observe that the stick and slip phases evolve differently with
$V$: the stick duration decreases with $V$, while the slip
duration  increases over the whole explored range of $V$. In
consequence, the ratio $T_{stick}/T_{slip}$, presented in
Fig.~\ref{rapports}, decreases with $V$ from $T_{stick}/T_{slip}
\sim 4 \pm 1$ down to $T_{stick}/T_{slip} \sim 0.3 \pm 0.2$. Such behavior of $T_{stick}/T_{slip}$ appears
to be very little dependent on $L$ according to
Fig.~\ref{rapports}. For $V \geqslant 0.90\pm 0.05$~m~s$^{-1}$,
$T_{stick}/T_{slip}$ becomes smaller than 1, meaning that the slip
phase is longer than the stick one. Our data therefore show that
it is not possible to neglect the slip duration compared to the
stick duration in general.

\section{Model}

In this section, we compare our experimental data with the model
proposed by Barquins, Maugis and co-workers in
refs.~\citenum{Barquins1986,Maugis1988}. This model is based on
measurements of the stable branch of the fracture energy
$\Gamma(v_p)$ for low peeling velocities below the instability
onset $V_a$, and on the following assumptions:
\begin{itemize}
\item During the stick phase, the equilibrium between the
instantaneous energy release rate $G=F/b$ and the fracture energy
$\Gamma(v_p)$ (of the low velocity stable branch) is still valid
dynamically, \textit{i.e.} $G(t)=\Gamma(v_p(t))$. \item The peeled
ribbon remains fully stretched during the peeling, which means
\begin{eqnarray}\label{eq:stretched}
G = \frac{F}{b} = \frac{E e}{L}\,u,
\end{eqnarray}
where $u$ is the elongation of the tape of Young modulus $E$ and
thickness $e$. \item The slip duration is negligible compared to
the stick duration.
\end{itemize}
Backing on these hypothesis, it is possible to derive a prediction
for the stick-slip duration $T_{ss}$. Introducing the inverted
function $v_p=\Gamma^{-1}(G)$ and noting that $du/dt=V-v_p$ (see next paragraph and ref.~\citenum{Cortet2013}),
eqn~(\ref{eq:stretched}) leads to the dynamical relation
\begin{eqnarray}
\frac{dG}{dt}=\frac{E e}{L}\,(V-\Gamma^{-1}(G)),
\end{eqnarray}
which can be integrated over the stick phase to get
\begin{eqnarray}\label{eq.tstick}
T_{stick}=\frac{L}{E e}\int_{G_0}^{G_a} \frac{dG}{V-\Gamma_{\rm
slow}^{-1}(G)}.
\end{eqnarray}
$G_a$ is the maximum value of $\Gamma (v_p)$ at the end of the
``slow'' stable branch $\Gamma_{\rm slow}(v_p)$. $G_0$ is the
minimum value of $\Gamma(v_p)$ at the beginning of the ``fast''
stable branch $\Gamma_{\rm fast}(v_p)$ (see Fig.~\ref{fig:force})
and is assumed to be also the value of $G$ at which the stick
phase starts on the slow branch after a slip phase.

In this model, the ribbon is assumed to remain taut during the
whole stick-slip cycle. In order to challenge the validity of this
hypothesis, let us estimate the evolution of the elongation $u(t)$
of the tape as a function of time. If we note $P(t)$ the peeling
point position and $M$ the point where the peeled tape is winded,
we can define the quantity $u(t)$ as the difference between the
distance $|\overline{MP(t)}|$ and the length of the peeled tape in
the unstrained state. If $u(t)$ is positive, this quantity indeed
measures the elongation of the tape as in
eqn~(\ref{eq:stretched}), whereas it measures the excess of slack
tape if it is negative. Following ref.~\citenum{Cortet2013}, one
can show that
\begin{eqnarray}\label{eq.ut}
u(t)=u_0+\int_0^t(V-v_p
(t))dt-\cos\theta\int_0^t(R\dot{\beta}-v_p(t))dt.
\end{eqnarray}
Since in our experiments the peeling angle $\theta$ is close to
$90\degree$ and the roller rotation velocity $R\,d\beta/dt$ sticks
to the imposed peeling velocity $V$ to a precision always better
than $\pm 1.5\%$,~\cite{Cortet2013} we finally have $u(t)\simeq
u_0+\int_0^t(V-v_p(t))dt$. The elongation/slack $u(t)$ increases
of $\Delta u = \int_{0}^{T_{stick}}(V-v_p(t))dt$ during the stick
phase and decreases of the same amplitude $\Delta u
=-\int_{T_{stick}}^{T_{ss}}(V-v_p(t))dt$ during the slip phase.
This compensation is ensured by the fact the averaged velocity
over the stick-slip cycle matches the imposed velocity $V$, i.e.
$\int_{0}^{T_{ss}}(V-v_p(t))dt=0$, and is valid whether or not the
tape remains always taut during the stick-slip cycle.

To test the relevance of the hypothesis of a tape always in
tension, one can actually compare the increase/decrease $\Delta u$ of the
quantity $u(t)$ during the stick/slip phase to the one predicted by the
quasistatic model of Barquins and co-workers
\begin{eqnarray}\label{eq.Deltau}
\Delta u_{\rm theo} = \frac{L}{Eeb} (F_a-F_0)=\frac{L}{Ee}
(G_a-G_0),
\end{eqnarray}
for an always taut tape. Throughout our data, the relative
discrepancy $(\Delta u_{\rm theo}-\Delta u)/\Delta u$ is typically
less than 15\% which confirms the relevance of the assumption of a
tape in tension during the whole stick-slip cycle.

An equivalent but more instructive way to test the model of
Barquins and co-workers is to integrate numerically
eqn~(\ref{eq.tstick}) and compare it to experimental measurements
of stick duration. To do so, we use the fit of the data of energy
release rate $G(V)$ of Fig.~\ref{fig:force}, i.e.
$G(V)=\Gamma_{\rm slow}(V)=a\,V^{n}$, with $n=0.146$ and $a=137$.
The value of $G_0$ is affected by a significant uncertainty in our
data. We will therefore use two different guesses corresponding to
the limit values introduced at page 3 (see $G_{0,1}$ and $G_{0,2}$
in Fig.~\ref{fig:force}). These values of $G_0$ correspond to two
limit values of the fracture velocity at the beginning of the
stick phase: $V_{0,1}=10^{-6}$~m~s$^{-1}$ measured in another
adhesive but with a close behavior,~\cite{Maugis1988} and
$V_{0,2}=6.3\times 10^{-5}$~m~s$^{-1}$ which is an upper limit for
$V_0$ according to the data of Fig.~\ref{fig:force}.

\begin{figure}
\centerline{\includegraphics[width=7cm]{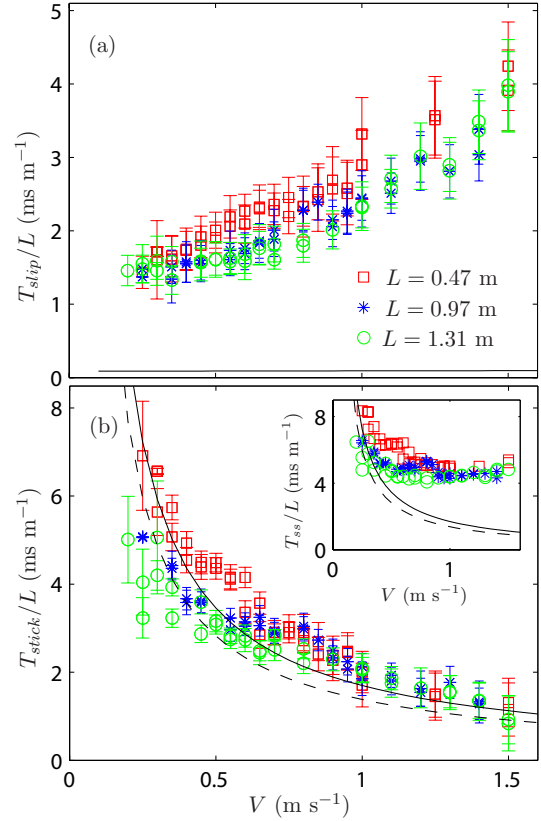}}
    \caption{\label{TsurL}(Color online) (a) $T_{slip}/L$, (b)
    $T_{stick}/L$ and  $T_{ss}/L$ (insert) \textit{vs.} $V$ for 3
    different $L$. Each data point corresponds to the average and each
    errorbar to the standard deviation of the statistics over one
    experiment. In (a), the curve close to the $x$-axis
    represents the theoretical prediction for a quasistationnary slip
    phase. In (b), the lines show the predictions of
    eqn~(\ref{eq.tstick}) with $V_a=0.10$~m~s$^{-1}$ and
    $V_{0,1}=10^{-6}$~m~s$^{-1}$ (solid line) or
    $V_{0,2}=6.3\times 10^{-5}$~m~s$^{-1}$ (dashed line).}
\end{figure}

In the insert of Fig.~\ref{TsurL}(b), we report the measured data
for $T_{ss}/L$ as a function of $V$ for three different lengths
$L$ as well as the predictions of eqn~(\ref{eq.tstick}) with
$V_{0,1}$ (solid line) and $V_{0,2}$ (dashed line). The model
appears compatible with the experimental data only for a marginal
range of  very low peeling velocities. Once $V>0.5$~m~s$^{-1}$,
the measured values of $T_{ss}/L$ indeed deviates more and more
from the theoretical prediction. A first natural explanation for
this discrepancy is that the assumption of a negligible slip
duration $T_{slip}$ (barely verified for low velocities for which
$0.25<T_{slip}/T_{stick}<0.5$) becomes more and more false as $V$
is increased (see Fig.~\ref{rapports}).

In Fig.~\ref{TsurL}(b) we therefore directly plot $T_{stick}/L$ as
a function of $V$, along with the prediction~(\ref{eq.tstick}).
One can note that the theoretical predictions using the two limit
guesses for $V_0$ are not very different. A first interesting
result is that the stick duration appears, to the first order,
proportional to the peeled tape length $L$ as evidenced by the
reasonable collapse of the data $T_{stick}/L$ on a master curve,
which is compatible with the analytical prediction of the
model~(\ref{eq.tstick}). But more importantly, we observe that for
the range of velocity explored, the model for $T_{stick}$, which do
not use any adjustable parameter, reproduces very well the
experimental data.

Obviously, one can consider an equivalent quasistationary
approximation during the slip phase in order to predict the slip
duration using $\Gamma_{\rm fast}^{-1}(G)$ instead of $\Gamma_{\rm
slow}^{-1}(G)$ in eqn~(\ref{eq.tstick}). Here, $\Gamma_{\rm
fast}^{-1}(G)$ corresponds to the inverse of the energy fracture
$G=\Gamma_{\rm fast}(v_p)$ in the fast and ``stable'' peeling
regime of Fig.~\ref{fig:force}. The integration using the model of the fast branch
$\Gamma_{\rm fast}(V)=6.5\times 10^{-5}\,V^{4.5}$  (see Fig.~\ref{fig:force}) however leads to values of $T_{slip}$
always 2 orders of magnitude smaller  than the experimental values
as evidenced in Fig.~\ref{TsurL}(a). It is however worth noting
that the collapse of the data $T_{slip}/L$ for the different $L$
shows that $T_{slip}$ increases nearly linearly with $L$.

\section{Discussion}

In this paper, we report experiments of a roller adhesive tape
peeled at a constant velocity focusing on the regime of stick-slip
instability. From fast imaging recordings, we extract the
dependencies of the stick and slip phases durations with the
imposed peeling velocity $V$ and peeled ribbon length $L$.

The stick phase duration $T_{stick}$ of the stick-slip
oscillations is shown to be nearly proportional to the peeled tape
length $L$ and to decrease with the peeling velocity $V$. These
data moreover appear in quantitative agreement with the
predictions of a model proposed by Barquins, Maugis and co-workers
in refs.~\citenum{Barquins1986,Maugis1988} which do not introduce
any adjustable parameter. This successful comparison confirms the
relevance of the two main assumptions made in the model: (i) the
tape remains in tension during the whole stick-slip cycle; (ii)
the principle of an equilibrium between the instantaneous energy
release rate $G(t)=F(t)/b$ and the fracture energy
$\Gamma(v_p(t))$, as measured in the steady peeling regime, is
valid dynamically during the stick phase.

Describing the peeling dynamics as a function of time $t$ by the
knowledge of the fracture velocity $v_p(t)$ and of the force
$F(t)=b\,G(t)$ in the peeled tape, the considered model further
assumes that the system jumps instantaneously, at the end of the
stick phase, from the ``slow'' stable branch to the ``fast''
stable branch of the steady fracture energy $G=\Gamma(v_p)$ and
then instantaneously backward from the ``fast'' branch to the
``slow'' branch at the end of the slip phase. In such a framework,
reproducing the assumptions (i) and (ii) for the slip phase leads
to a prediction for the slip duration. We have shown
that this prediction is at least hundred times smaller than the
slip phase duration $T_{slip}$ measured in our experiments. We
actually report that, contrary to what is finally proposed in
refs.~\citenum{Barquins1986,Maugis1988}, the slip duration
$T_{slip}$ cannot be neglected compared to the stick one
$T_{stick}$, since it is at best 4 times smaller, and becomes even
larger than $T_{stick}$ for $V \geqslant 0.90 \pm
0.05$~m~s$^{-1}$.

These last experimental results account for the existence of
strong dynamical effects during the slip phase which can therefore
not be described by a quasistatic hypothesis. These dynamical
effects could be due to the inertia of the ribbon close to the
fracture front. Some models also predict a strong
influence of the roller inertia.\cite{Ciccotti2004, De2008}
Notably, thanks to numerical computation, De and
Ananthakrishna\cite{De2008} have shown that for certain values of
the roller inertia, the slip phase could consist in several jumps
from the ``fast stable'' branch to the ``slow stable'' branch in
the $(v_p,G=\Gamma(v_p))$ diagram. Such a process would certainly produce a longer slip time than expected in the framework of Barquins's model. It would be most interesting to
confront our experimental observations to the predictions of this
model, based on a detailed set of dynamical equations and ad-hoc
assumptions made on the velocity dependence of $\Gamma$. However, such a comparison is not straightforward in our current setup since we do not have the temporal and spatial resolutions to detect such eventual fast oscillations. Besides, in order to obtain a quantitative comparison, measurement of the instantaneous peeling force $F(t)$ is required but it remains a challenge.

\section*{Acknowledgments}

This work has been supported by the French ANR through Grant
``STICKSLIP'' No. 12-BS09-014. We thank Costantino Creton and
Matteo Ciccotti for fruitful discussions and Matteo Ciccotti for
sharing data with us.

\end{document}